%% file: nbc_paper.tex
\documentclass[useAMS,usenatbib,letterpaper]{mn2e}

\usepackage{amsmath,mathtools,graphicx,amssymb,verbatim,url,epsfig,natbib,bm,psfrag,ifthen,hyperref}
\usepackage[english]{babel}
\usepackage{epsfig}
\usepackage{times}

\renewcommand{\vec}[1]{\mathbf{#1}}
\providecommand{\e}[1]{\ensuremath{\cdot 10^{#1}}} 

\providecommand{\mean}[1]{\ensuremath{\langle #1 \rangle}}

\title[Calibration of noise bias for weak lensing]{Measurement and Calibration of Noise Bias in Weak Lensing Galaxy Shape Estimation}
\author[Tomasz Kacprzak et al.]{
Tomasz~Kacprzak,$^{1}$\thanks{E-mail: \texttt{tomasz.kacprzak.09@ucl.ac.uk}}
Joe~Zuntz,$^{1,2,3}$
Barnaby~Rowe,$^{1,4,5}$
Sarah~Bridle,$^{1}$\newauthor
Alexandre~Refregier,$^{6}$
Adam~Amara,$^{6}$
Lisa~Voigt,$^{1}$
Michael~Hirsch$^{1,7}$ 
\\
$^{1}$Department of Physics \& Astronomy, University College London, Gower Street, London WC1E 6BT\\
$^{2}$Astrophysics Group, University of Oxford, Denys Wilkinson Building, Keble Road, Oxford OX1 3RH\\
$^{3}$Oxford Martin School, University of Oxford, Old Indian Institute, 34 Broad Street, Oxford OX1 3BD\\
$^{4}$Jet Propulsion Laboratory, California Institute of Technology,\thanks{This
  work was not done in the author's capacity as an employee of the Jet
  Propulsion Laboratory, California Institute of Technology.} 4800 Oak Grove Drive, Pasadena, CA 91109, USA\\
$^{5}$California Institute of Technology, 1200 East California Boulevard, Pasadena, CA 91125, USA\\
$^{6}$Institute for Astronomy, Eidgen\"{o}ssische Technische Hochschule Zurich,
Wolfgang-Pauli-Stasse 27, CH-8093 Zurich, Switzerland\\
$^{7}$Max Planck Institute for Intelligent Systems, Department of Empirical Inference, Spemannstra\ss{}e 38, 72076 T\"ubingen, Germany \\
}

\begin{document}

\date{\today}

\pagerange{\pageref{firstpage}--\pageref{lastpage}} \pubyear{2011}

\maketitle

\label{firstpage}

\begin{abstract}
Weak gravitational lensing has the potential to constrain cosmological parameters to high precision. 
However, as shown by the Shear TEsting Programmes (STEP) and GRavitational lEnsing Accuracy Testing (GREAT) Challenges, measuring galaxy shears is a nontrivial task: various methods introduce different systematic biases which have to be accounted for.
We investigate how pixel noise on the image affects the bias on shear estimates from a Maximum-Likelihood forward model-fitting approach using a sum of co-elliptical S\'{e}rsic profiles, in complement to the theoretical approach of an an associated paper.
We evaluate the bias using a simple but realistic galaxy model and find that the effects of noise alone can cause biases of order 1-10\% on measured shears, which is significant for current and future lensing surveys.  
We evaluate a simulation-based calibration method to create a bias model as a function of galaxy properties and observing conditions. 
This model is then used to correct the simulated measurements. 
We demonstrate that this method can effectively reduce noise bias so that shear measurement reaches the level of accuracy required for estimating cosmic shear in upcoming lensing surveys.
\end{abstract}


\input{nbc_paper_introduction}

\input{nbc_paper_shear}

\input{nbc_paper_evaluation}

\input{nbc_paper_calibration}
\input{nbc_paper_summary}

\input{nbc_paper_references}
\input{nbc_paper_appendix1}

\end{document}

%% file: nbc_paper_introduction.tex
\section{Introduction}
\label{sec:Introduction} 

Weak gravitational lensing is an important cosmological probe, which has the greatest potential to discover the cause of the accelerated cosmic expansion \citep[e.g.][]{esoesa,detfr2006,albrecht2009}. 
In the standard cosmological model dark energy affects both the expansion history of the universe and the rate of gravitational collapse of large scale structure. 
The rate of this collapse can be studied by observing the spatial distribution of dark matter at different times in the history of the universe. 
Gravitational lensing occurs when the path of light from distant galaxies is perturbed while passing through intervening matter. 
This phenomenon causes the images of galaxies to be distorted. 
The primary observable distortion is called gravitational shear, and typically causes the galaxy images to be stretched by a few percent. 
The scale of this effect is related to the amount of matter between the source and the observer, and to their relative geometry. 
Thus, cosmic shear can provide a valuable dataset for testing cosmology models \citep{Kaiser1992,hu99}.

Several upcoming imaging surveys plan to observe cosmic shear, including the 
KIlo-Degree Survey: KIDS, 
the Dark Energy Survey (DES)\footnote{http://www.darkenergysurvey.org},
the Hyper Suprime-Cam (HSC) survey\footnote{http://www.naoj.org/Projects/HSC/HSCProject.html}
%S4 \footnote{http://www.naoj.org/Projects/HSC/index.html}, %S3 [not sure how to cite this properly - I emailed Rachel M.]
%JAZ I added a possibe URL so we do not forget if Rachel does not get back to us
%S4 [Thanks Joe. Now fixed ref according to reply from Rachel Mandelbaum]
the Large Synoptic Survey Telescope (LSST)\footnote{http://www.lsst.org}, 
Euclid\footnote{http://sci.esa.int/euclid} 
and WFIRST~\footnote{http://exep.jpl.nasa.gov/programElements/wfirst/}. 
For these surveys, it is crucial that the systematics introduced by data analysis pipelines are understood and accounted for. 
The most significant systematic errors are introduced by 
(i) the measurements of the distance to the observed galaxies using photometric redshifts, 
(ii) intrinsic alignments of galaxies, 
(iii) modelling of the clustering of matter on the small scales in the presence of baryons, 
(iv) measurement of lensed galaxy shapes from imperfect images. 
In this paper, we focus on the latter. 

To evaluate the performance of shear measurement methods, simulated datasets have been created and released in form of blind challenges. 
The Shear TEsting Programme 1 (STEP1: \citealp{step1}), was the first in this series, followed by STEP2 \citep{step2}. 
Both challenges aimed to test end-to-end shear pipelines and simulated galaxy images containing many physical effects including those stemming fom telescope optics and atmospheric turbulence. 
A modified approach was taken in the GREAT08 \citep{great08handbook,great08results} and GREAT10 \citep{great10handbook} challenges, which sought to isolate independent parts of the data analysis process. 
They explored the impact of different true galaxy and image parameters on the shear measurement, by varying them one at a time among various simulation realisations. 
These parameters included signal to noise ratio, galaxy size, galaxy model, Point Spread Function (PSF) characteristics and others. 
The results showed that the shear measurement problem is intricate and complex. 
Existing methods proved to be sufficient for current surveys, but there is room for improvement for the future. 

For a well resolved, blur-free, noise-free image, the galaxy ellipticity can be calculated by taking the moments of the image \citep{BonnetMellier1995}. 
However, a typical galaxy image used in weak lensing is highly affected by the observation process. 
The image degrading effects are 
(i) convolution with the PSF of the telescope, 
(ii) pixelisation of the image by the light buckets of the detector, 
(iii) pixel noise on the image due to the finite number of photons from the source and atmosphere (roughly Poisson) and detector noise (often assumed Gaussian), and 
(iv) galaxy colours being different from the stars used to map the PSF \citep{cypriano2010} and a function of position on the galaxy \citep{Voigt2011}.

Moment-based methods such as KSB \citep{ksb1}, and most recently DEIMOS \citep{deimos1}, and FDNT \citep{bernstein2010} measure the quadrupole moment of the image, 
using a masking function (often Gaussian) to counter the effects of noise, and then correct for the PSF. 
Decomposition methods, e.g. shapelets or a Gauss-Laguerre expansion, \citep{shapelets1,BernsteinJarvis2002,GL1} use an orthogonal image basis set which can be easily convolved with the PSF. 
Noise is accounted for by regularisation of the coefficients matrix and truncating the basis set to a finite number of elements. 
Simple model fitting methods based on sums of Gaussians \citep{Kiijken1999, im2shape}, S\'{e}rsic profiles \citep{lensfit1, galfit}, create an ellipticity estimator from a likelihood function. 
Stacking methods \citep{lewis2009,hosseini2009}, which have been demonstrated for constant shear fields, average a function of the image pixels to increase the signal-to-noise ratio and then deconvolve the PSF.

All these methods introduce some level of systematic error. 
Bias on the shear can result from inaccurate centroiding of the galaxy, for example see \citet{lewis2009}. 
Another source, model bias, results from using a galaxy model which does not span the true range of galaxy shapes.
\citet{voigt2010} quantified the shear measurement bias from using an elliptical isophote galaxy model on a galaxy with a a more complicated morphological structure in the presence of a PSF \citep[see][for a general proof]{lewis2009}.  
\citet{melchior2010} investigated the effectiveness of shapelets at representing more realistic galaxies. 
\citet{viola2011} and \citet{bartelmann2011} quantified biases on the KSB method and investigated possibilities to correct for it. 
 
Pixel noise bias arises from the fact that ellipticity is not a linear function of pixel intensities in the presence of noise and PSF.
\citet{hirataetal04} showed its effects on second order moment measurements from convolved Gaussian galaxy images. 
The bias due to pixel noise on parameters fitted using Maximum Likelihood Estimators (MLEs) for elliptical shapes was demonstrated by \citep[][hereafter R12]{NoiseBias1}, for the case when the noise is Gaussian and the correct galaxy model is known. 
It presented a general expression for the dependency of the bias on the signal to noise ratio. 
It also demonstrated the consistency of analytical and simulated results for the bias on the width for a one parameter Gaussian galaxy model. 
%S3 [I'm commenting out the next two sentences as they are too cryptic to be useful, 
%S3 and too interesting to leave unjustified. 
%S3 e.g. who believes this? why is this believed? which other estimators were considered?
%S3 What do you have in mind for the second sentence? GREAT08 results? 
%S3 which methods? would have to go through each and justify the statement I think.
%S3 Also if we talk about more general biases/work here then the next paragraph will need re-writing as it currently refers to Alex's paper]]
%S3 Although the maximum likelihood was used, it is believed that the noise bias will also exist for other estimators. Hitherto no unbiased estimators of shape in the presence of a PSF were demonstrated. 

In this paper, we show the significance of this bias for weak lensing measurements using more realistic galaxy images. 
We find that the bias as a function of true input parameters is consistent with the theoretical framework derived in R12.
Furthermore, we present a method to effectively remove this noise bias for realistic galaxy images. 
Using the \textsc{Im3shape} shear measurement framework and code (Zuntz et al. in prep), we use a forward model fitting, Maximum Likelihood (ML) approach for parameter estimation. 
We create a model of the bias as a function of galaxy and PSF parameters by determining their bias from various simulations that sample parameter space. 
We apply this model to the noisy MLEs and demonstrate that this procedure successfully removes the noise bias to the accuracy required by upcoming galaxy surveys. 
By performing a calibration that depends on the specific statistics of every recorded galaxy, this method is independent of the overall galaxy and PSF parameter distributions.

This paper is organised as follows. Section \ref{sec:Shear} summarises the equations governing the cosmic shear measurement problem and describes methods to quantify the biases on estimated parameters. 
We also discuss the requirements on those biases for lensing surveys, followed by a summary of the cause of bias arising from image noise.
In Section \ref{sec:Evaluation}, we show the results of bias measurements. 
A method for correcting the noise bias based on numerical simulations is presented in Section \ref{sec:Application}.
We conclude and briefly discuss this approach and alternatives in Section  \ref{sec:Wrap}. 
In the Appendices, we detail the method used for measuring the multiplicative and additive bias and tabulate our results and fit parameters.

%% file: nbc_paper_shear.tex
\section{Shear measurement biases in model fitting} %S1
\label{sec:Shear}
We first discuss the parametrisation of shear measurement biases, and present an overview of the model fitting approach. 
We summarise recent work on noise bias in a simple case, and then describe our shear measurement procedure and simulation parameters. 

\subsection{Quantifying systematic biases in shear estimation} %S1
\label{sec:defns_systematics}

In weak gravitational lensing the galaxy image is distorted by a Jacobian matrix \citep[see][for reviews]{BartelmannSchneider2001,BernsteinJarvis2002,HoekstraJain2008}
\begin{eqnarray}
\emph{\textbf{M}}  
%~ = ~ \frac{\partial\bbeta}{\partial\btheta}
% & = & \left[
%                    \begin{array}{cc}
%                      1-\psi_{11} & -\psi_{12} \\
%                      -\psi_{21} & 1-\psi_{22} \\                      
%                    \end{array}
%                  \right] \\
& =& \left[
                    \begin{array}{cc}
                      1-\kappa-\gamma_{1} & -\gamma_{2} \\
                      -\gamma_{2} & 1-\kappa+\gamma_{1} \\
                    \end{array}
                  \right],
\end{eqnarray}
where $\kappa$ is the convergence and 
$\gamma=\gamma_{1}+i\gamma_{2}$
is the complex gravitational shear.

For a galaxy with elliptical isophotes we can define the complex ellipticity $e$ as 
\begin{equation}
	e=\frac{a-b}{a+b}\,\,e^{2i\phi}, %S3
\end{equation}
where $b/a$ is the galaxy minor to major axis ratio and $\phi$ is the orientation of the major axis anticlockwise from the positive $x$-axis.
The post-shear lensed ellipticity $e^l$ is related to the intrinsic ellipticity $e^i$ by 
\begin{equation}
e^{l}=\frac{e^i+g}{1+g^{*}e^{i}}
\label{eqn:eobs}
\end{equation}
for $|g|\leq1$ %L1
\citep{seitzs97}, where
$g=\gamma/(1-\kappa)$
is the reduced shear.
In the weak lensing regime $\kappa\ll1$, $\gamma\ll1$ and $g\approx\gamma$. 
We assume $\kappa\ll1$ throughout this paper. %L1

Galaxies have intrinsic ellipticities which are typically an order of magnitude larger than the shear. 
For a constant shear and an infinite number of randomly orientated galaxies the mean lensed ellipticity is equal to the shear, to third order in the shear. 
In practice $e^l$ is averaged over a finite number of galaxies and the error on the shear estimate (referred to as `shape noise') depends on the distribution of galaxy intrinsic ellipticities and the number of galaxies analysed.     

The accuracy of a shape measurement method can be tested on a finite number of images in the absence of shape noise by performing a `ring-test' \citep{GL1}.
In the ring-test, the shear estimate is obtained by averaging the measured $e^o$ estimates from a finite number of instances of a galaxy rotated through angles distributed uniformly from 0 to 180 degrees.
If $\hat e^{l}$ is the measured lensed ellipticity, then the shear estimate is $\hat \gamma = \langle \hat e^{l} \rangle$ and the bias on the shear is
\begin{equation}
 b[\hat \gamma] = \langle \hat e^{l} \rangle- 
 %L1\gamma^{true}.
 \gamma^{t},
\label{eqn:bias_on_shear}
\end{equation}
where $\gamma^{t}$ is the true shear.
This bias on the shear is usually quantified in terms of multiplicative and additive errors $m_{i}$ and $c_{i}$ for both shear components $i={1,2}$ such that 
\begin{equation}
\hat{\gamma}_i=(1+m_i) \gamma^{t}_i+c_i, %L1 [changed j to i, true to t]
\end{equation}
assuming $\hat \gamma_1$ does not depend on $\gamma_2^{t}$, and vice versa \citep[][]{step1}. 
The requirements on the level of systematic errors for current and future galaxy surveys are expressed in terms of $m_i, c_i$ in \cite{amara2008} and are summarised in Table \ref{tab:requirements}.
%S3 [actually, do we ever refer to anything other than "upcoming" surveys in the text? if not then I suggest removing the table andjust writing about upcoming surveys in the text. But perhaps we should make sure to discuss far future too]
%S4 [we do now discuss far future in the text]
\begin{table}
\center
\begin{tabular}{|c|ccc|ccc|}
\hline
Survey   &  $m_i$ & $c_i$  \\
\hline
Current  & 0.02  & 0.001   \\
Upcoming future & 0.004 & 0.0006  \\ 
Far future   & 0.001 & 0.0003  \\ 
\hline
\end{tabular}
\caption{Summary of the requirements for the bias on the shear for current, upcoming and far future surveys.}
\label{tab:requirements}
\end{table}

\subsection{Galaxy shear from model fitting} %S1

A simple approach to measuring ellipticity is to use a parametric model. 
For galaxy fitting, models such as sums of Gaussians \citep{Kiijken1999,im2shape}, S\'{e}rsic profiles \citep{lensfit1}, and Gauss - Laguerre polynomials (shapelets) \citep{shapelets1, BernsteinJarvis2002, GL1} were used. 

In general, model fitting methods are based on a likelihood function. 
Under uncorrelated Gaussian noise, this function is
\begin{align}
\mathcal{L} &= p(\vec{\theta} | \vec{I}, \mathcal{M})  \\
\log \mathcal{L} &= \chi^2 = \frac{1}{2} \sum_{i=1}^N [M_i(\vec{\theta}) - I_i]^2 / \sigma_i^2
\label{eqn:chi2} 
\end{align}
where $\vec{\theta}$ is a set of variable model parameters, $\vec{I}$
is the observed galaxy image, $\mathcal{M}$ is a model function,
$\vec{M}(\vec{\theta})$ is the model image created with parameters
$\vec{\theta}$, and $N$ the number of pixels in images $\vec{I}$ and $\vec{M}$. 
These equations assume a known noise level on each pixel $\sigma_i$, which is often assumed constant $\sigma_i = \sigma_{\rm noise}$. 
%S3 [NB if we let the noise level float then there is an extra term in the log likelihood - we could consider doing this.]
Sometimes a prior on the parameters is used to create a posterior function. 

Usually an ellipticity estimator is derived from this likelihood function; so far maximum likelihood estimators (MLE; e.g. Im3shape, Shapelets), mean likelihood (Im2shape) and mean posterior (e.g. LensFit) have been used. 
We use the MLE in this paper. 

Parametric models based on elliptical profiles typically use the following galaxy parameters: centroid, ellipticity, size, flux and a galaxy light profile parameter.
Often a combination of two S\'{e}rsic profiles \citep{sersic63} is used to represent the galaxy bulge and disc components, with identical centroids and ellipticities. 

The model also contains information about other effects influencing the creation of the image.
These image parameters are not often a subject of optimisation: noise level $\sigma_{\rm noise}$, PSF kernel and the pixel integration kernel. 
SNR is often defined as
${\rm SNR} = \sqrt{\sum_{i=1}^N I_i^2} / {\sigma_{\rm noise}}$
and this definition will be used throughout this paper. 
This definition of SNR is the same as in GREAT08, but different to GREAT10: SNR=20 here corresponds to SNR=10 in GREAT10.

\subsection{Noise bias} 
\label{sec:NoiseBias}
%JAZ What about the Hirata paper?
The bias of parameter estimation for MLEs in the context of galaxy fitting was first studied by R12. 
The authors derived general expressions for the covariance and bias of the MLE of a 2D Gaussian galaxy model convolved with a Gaussian PSF. 
For a nonlinear model, in the Taylor expansion of $\chi^2$ (in equation \ref{eqn:chi2}) the terms in even power of the noise standard deviation are found to contribute to the the estimator bias.
The analytical results were confirmed by simulations using a single parameter toy model. 
It was also noted that the bias is sensitive to the chosen parametrisation, especially if the parameter space is bounded.   
 
\subsection{\textsc{Im3shape} pipeline} 
\label{sec:im3shape}
The analyses in this paper were performed using the \textsc{Im3shape} shear measurement framework and code.  
Here we outline the system,  which will be described in more detail in Zuntz et al. (in prep). 

Each simulated galaxy is fitted with a model containing two co-centric, co-elliptical S\'{e}rsic components, one de Vaucouleurs bulge (S\'{e}rsic index=4) and one exponential disc (S\'{e}rsic index=1).  
The amplitudes of the bulge and disc were free but the ratio of the 
half light %S3 [need to say what sort of radii we are talking about. I'm guessing its the scale radii not the FWHM?]
radii was fixed to 
%TK {\bf 1.23}.  % Barney: agree with Sarah about the scale radii, ambiguous.  Also think we need a
%ref for Sersic profiles...
1.0. 
They are convolved with the true Moffat PSF model to produce a model image. 
Since there is high resolution structure in de Vaucouleurs bulges we made the models at a higher resolution than the final images. 
We use a resolution three times higher in the outer regions and 45 times higher in the central 3$\times$3 pixels of the final image.
Since very highly elliptical images are hard to simulate accurately we restrict the allowed space of models to those with $|e|<0.95$. 

We find the peak of the likelihood using the Levenberg-Marquadt method \citep{Lourakis} using numerical gradients of each image pixel in the likelihood. 
We tested the performance of the optimiser for variety of input galaxy and image parameters to ensure that the optimiser always converges to a local minimum by evaluating the likelihood in the neighbourhood of the found best fit point for multiple test noise realisations.
In this nonlinear optimisation problem multiple likelihood modes are possible. 
However, for our simple model, we found that usually there was only one local minimum (i.e. the bias results did not depend on the starting parameters given to the minimiser). 
We will discuss this further in Zuntz et al. (in prep). 

\subsection{Simulation parameters} %S1
\label{sec:defaults}

The galaxies used for this study were created using a two component model: 
a S\'{e}rsic profile of index 4 for the bulge and a S\'{e}rsic profile of index 1 for the disc. 
Both components have the same centroid, ellipticity and scale radius. 
%S3 They also have the same centroid. 
%S3 Although this model may not be realistic enough for galaxy surveys, we believe that the procedure should be possible to repeat with a model including more parameters. 
%S3 [er, I don't believe this! Lets defer this discussion to the conclusions section though]
The galaxy model used for fitting encompassed the one used to create the true galaxy image; therefore we are isolating the noise bias effect from the model bias effect in this study. 
The PSF was modelled as a Moffat profile with a FWHM of 2.85 pixels
and Moffat $\beta$ parameter of 3 (see, e.g., \citealp{great08results}
for a definition of the Moffat and the notation adopted here). 
We use the same PSF in the fit as in the simulated images to prevent
any bias effects caused by incorrect modelling of the PSF. 
We fit a total of 7 parameters: galaxy centroid $x$, $y$; galaxy ellipticity $e_1$, $e_2$; galaxy size $r$; bulge flux $F_b$; and disc flux $F_d$. 

We expect variation in the following physical parameters to have the most significant influence on the noise bias, and therefore the bias will be evaluated as a function of:
\begin{itemize}
\item 
Signal-to-noise ratio (SNR),
\item 
Intrinsic galaxy ellipticity,
\item 
PSF ellipticity,
\item 
Size of the galaxy compared to the size of the PSF, expressed as $R_{gp}/R_{p}$, which  is the ratio of the FWHM of the convolved observed object and the FWHM of the PSF. 
Note that this is not the same as the parameter we fit. This is because the noise bias strongly depends on the PSF parameters, and the galaxy radius parameter alone would not fully capture this dependence.
\item 
Light profile of the galaxy, described by the flux ratio $F_b/(F_b+F_d)$, which is the flux of the bulge component divided by total flux of the galaxy.
For a purely bulge galaxy, $F_b/(F_b+F_d)= 1$ and for a disc galaxy $F_b/(F_b+F_d)=0$.
In our model, we allow the amplitudes of the components to be negative, so the flux ratio can take both values $F_b/(F_b+F_d)>1$ and $F_b/(F_b+F_d)<0$. 
Therefore, for $F_b/(F_b+F_d)>1$, the galaxy has a negative disc component, which results in the galaxy being less `peaky' than a galaxy with $F_b/(F_b+F_d)=1$, and the galaxy model image may even be more similar to a galaxy with $F_b/(F_b+F_d)<1$. 
An alternative might be to use a more flexible radial profile, for example a larger number of Sersic components, or allowing the Sersic indices to be free parameters in the fit. 
\end{itemize}

These parameters will be used to create a model for the noise bias. 
We expect these physical parameters to best encapsulate the main dependencies of the bias, although we are aware that there may exist other statistics that better capture bias variation.  
 
We do not show the effect of the galaxy centroid on the bias, as no
significant dependence on this parameter was found in our experiments. 
We measured the noise bias for a simulated galaxy image with identical
model parameters, once located in the middle of a pixel and once on the edge of a pixel. 
We found no difference in ellipticity bias to our desired precision. 
In the simulations, the galaxy centroid is randomised. 
 
The values for the simulation parameters are summarised in Table \ref{tab:SimParams}. 
Their choice is based on galaxies used in GREAT08.
We define a fiducial parameter set and make departures D1 to D4 in one parameter at a time using the values given in the Table.
We restrict our analysis to SNR values of 20 and greater because we
find convergence of the minimiser does not pass our quality tests at
lower values. 
%MH [Maybe we should describe what these quality tests are?]
However, the SNR values of most interest for upcoming surveys are low, and therefore we use the lowest SNR we can use with confidence by default for all simulations. 
We investigate a SNR value of 200 which matches that of the GREAT08 LowNoise simulation set, plus an intermediate value of 40 which is also used in GREAT08.
By default, we use a galaxy with half the flux in a bulge and half in a disc. 
The two perturbations we consider are to pure bulge and pure disc. 
Finally we explore the dependence of noise bias on the PSF ellipticity, spanning the range from zero to 10\%.

For the minimisation parameters used in this paper,  \textsc{Im3shape} takes around one second per galaxy, which is typical for model fitting methods.
%S4 [could cite GREAT08 appendix here]
To obtain our desired accuracy on noise bias we needed to simulate 2.5 million galaxies for each set of simulation parameters shown in Table \ref{tab:SimParams}.
Therefore the computations shown in this paper took of order 1 year of CPU time.
This computational burden limited the number of points we could show on the figures to 3 per varied parameter.

\begin{table}
\center
\begin{tabular}{| c | c |c |c|}
\hline
		& Parameter 		&  Fiducial & Deviations \\
\hline
D1		& 
SNR 		& 20 	& 40, 200   \\ 
D2		& $R_{gp}/R_{p} 	$ & 1.62 	& 1.41, 1.82   \\ 
D3		& $F_{b}/(F_{b} + F_{d})$ 	& 0.5 	& 0 , 1   \\    
D4		& $e_1^{PSF}$ 			& 0.05 	& 0, 0.1  \\ \hline
\end{tabular}
\caption{Summary of parameters used for simulations. For D3 two different parametrisations are shown for clarity.}
\label{tab:SimParams}
\end{table}

%% file: nbc_paper_evaluation.tex
\begin{center}
\begin{figure*}
\epsfig{file=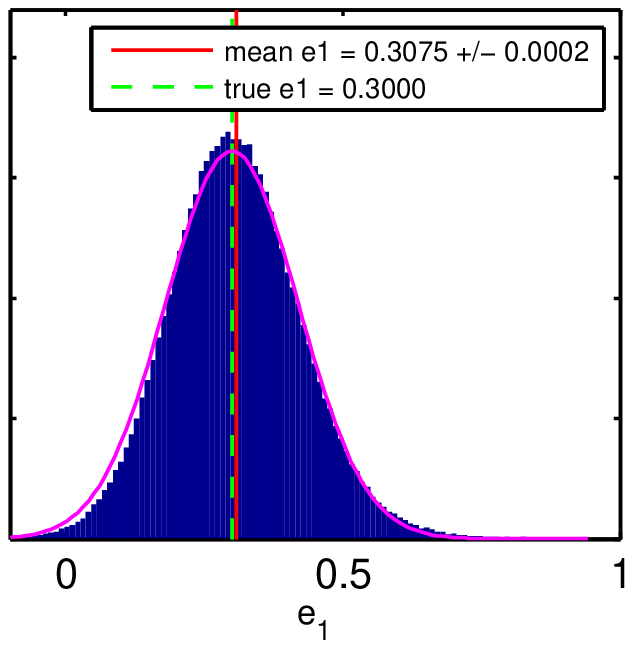,width=8cm,height=7cm} 
\epsfig{file=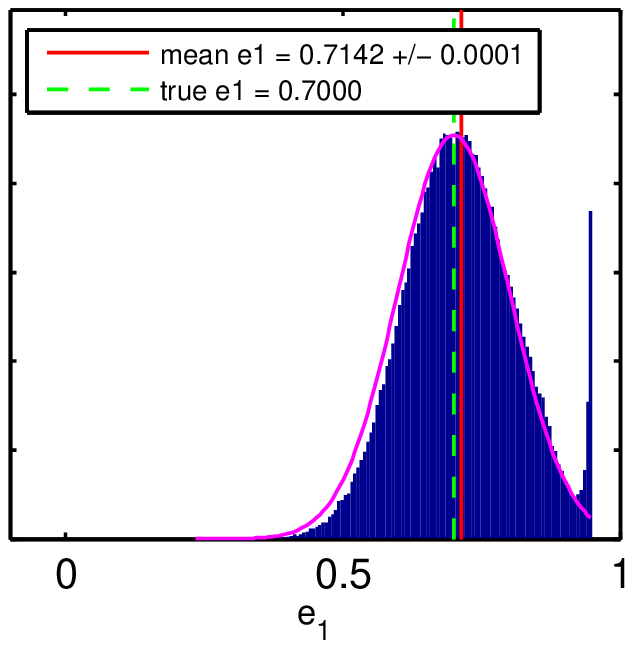,width=8cm,height=7cm}
\epsfig{file=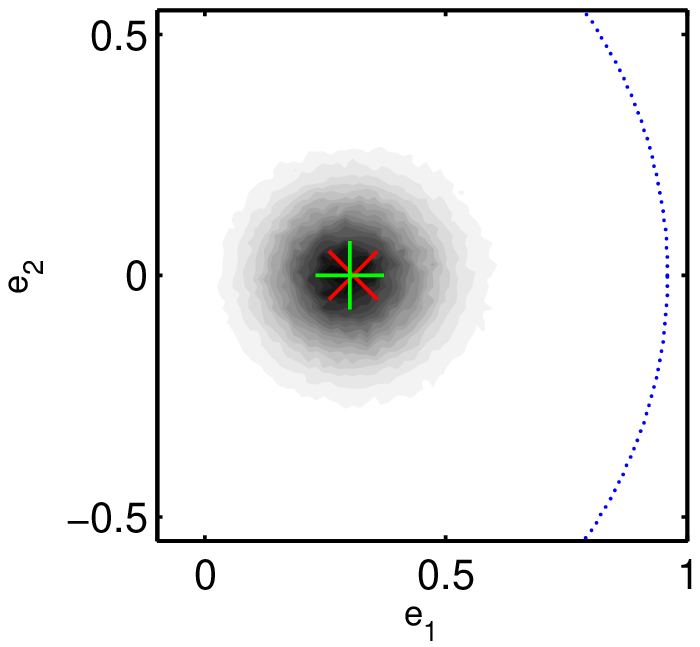,width=8cm,height=7cm}
\epsfig{file=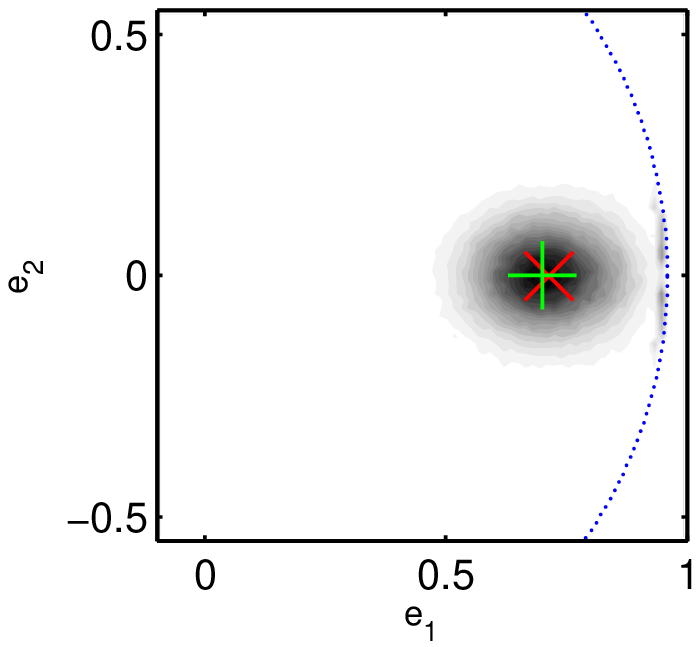,width=8cm,height=7cm}
\epsfig{file=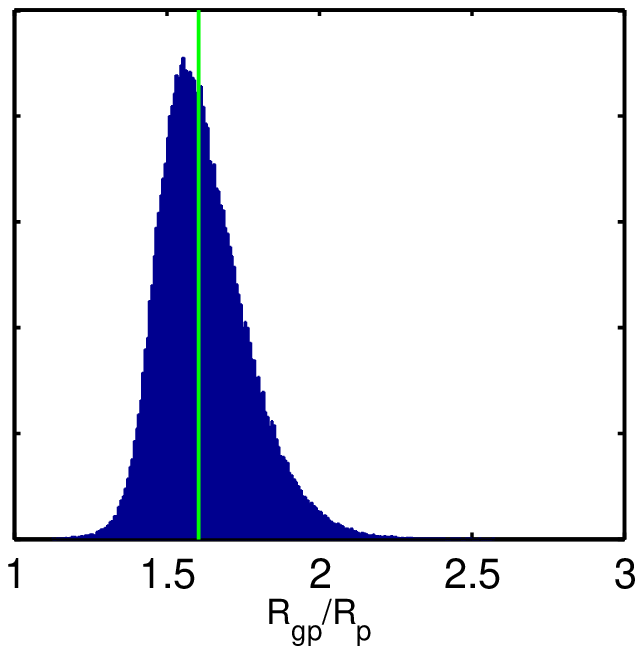,width=8cm,height=7cm}
\epsfig{file=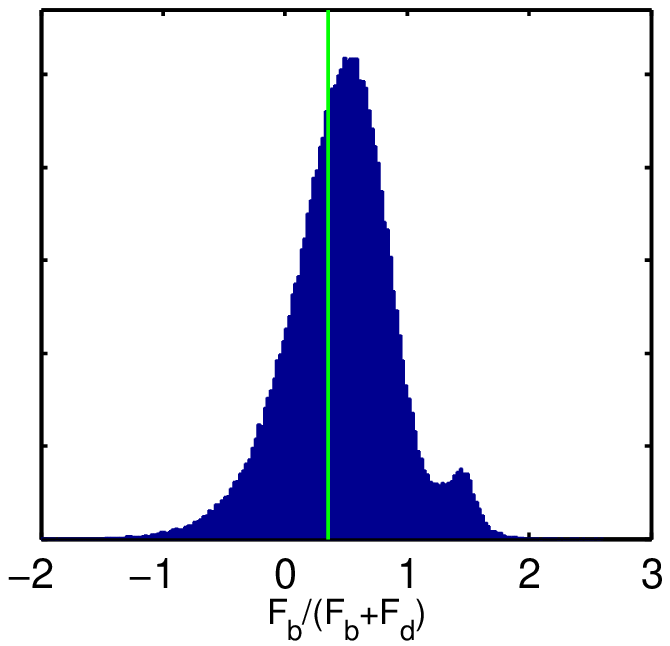,width=8cm,height=7cm}
\caption{
Histograms of ML parameter estimates for the fiducial galaxy model. 
Top panels show the distribution of measured ellipticity ($e_1$) parameters for true intrinsic galaxy ellipticity of 0.3 (left) and 0.7 (right), marked with green dashed line. 
The empirical mean of these distributions is marked with a red solid line. 
The middle panels shows the distribution of the ML estimates for both ellipticity components -- for true intrinsic ellipticity of [0.3,0.0] (left) and [0.7,0.0] (right), marked with the plus sign.  
The mean of this distribution is marked with a cross sign. 
The effective boundary on the ellipticity parameter space ($|e|=0.95$) is marked with black dotted line. 
The bottom panels show histograms of measured size ($R_{gp}/R_{p}$) and light profile ($F_b/(F_b+F_d)$) parameters. 
True values for these parameters are marked with red solid line -- true $R_{gp}/R_{p}= 1.6$ and true $F_b/(F_b+F_d)$.
}
\label{fig:histograms} 
\end{figure*}
\end{center} 

\section{Evaluation of the noise bias effect} %S1
\label{sec:Evaluation}

In this section we evaluate the noise bias as a function of galaxy and image parameters.
We define the noise bias on an ellipticity measurement as 
\begin{equation}
b[\hat e] = \langle \hat e \rangle - e^{\rm true}.
\end{equation} 
We calculate the bias using the following procedure: 
we create a galaxy image with some true ellipticity, add a noise map and measure the MLE of the ellipticity. 
Then, we repeat this procedure with different noise realisations which results in a distribution of noisy MLE ellipticities.
The difference between the mean of this distribution and the true galaxy ellipticity is the bias on ellipticity. 

The histograms of ML estimates for 300 thousand noise realisations are plotted in Figure \ref{fig:histograms} to illustrate the nature of the noise bias. 
The galaxy and image had default parameters described in Section \ref{sec:defaults} and intrinsic ellipticities of $e_1=0.3$, $e_2=0$ and $e_1=0.7$, $e_2=0$ in the left and right upper and middle panels, respectively. 
The spread of values comes from the Gaussian noise added to the images to approximate the finite number of photons arriving on the detector. 
As discussed in Section~\ref{sec:defaults}, we assume a default SNR value of 20.

Two effects contribute significantly to the bias on ellipticity for the left hand panels in which the true ellipticity is $e_1=0.3$, $e_2=0$.  
The ellipticity distribution is slightly skewed away from being a Gaussian. 
There is a larger tail to high ellipticity values than to negative ellipticity values. 
The peak is shifted to lower ellipticities, which is also visible in the two-dimensional histogram in the middle-left panel of Figure~\ref{fig:histograms}. 
%S3 [help! to be honest I can't see the peak shift]
%S3 [we can't say in the next sentence "this net positive bias" without describing the net positive
%bias first. Added a sentence on this:]
% Barney: I really can't see the peak shift!
%S4 !!! OK, thanks for confirming Barney. I still can't see it either! Am modifying the text
%S4 [maybe we can change the figures somehow to make it visible??]
Overall there is a net positive bias to larger ellipticity values, as shown by the vertical solid line which is to be compared with the vertical dashed line placed at the true value. 
Although this net positive bias is hard to see by eye, it is significant at the level of shear measurement accuracy required from future observations.
This is discussed in more detail in the following sections.  

Furthermore, the ellipticity parameter space is theoretically bounded at an ellipticity modulus of unity. 
This is exacerbated by any realistic measurement method which will break down just short of unity. 
The consequence of this effect is visible for a galaxy with true intrinsic ellipticity of $|e|=0.7$, shown in the upper-right and middle-right panels. 
For this example, it counteracts the noise bias effect by reducing the amount of overestimation. 
For more noisy or smaller galaxies, which will have larger variance in the ellipticity MLEs, this effect will be stronger and may even cause the ellipticity to be underestimated, see \ref{fig:polyfit} for an illustration of this. 

Distributions of other fitted parameters are also biased and skewed, as discussed in R12.
We show histograms of fitted galaxy size and galaxy light profile in
the two bottom panels of Figure~\ref{fig:histograms}.
The convolved galaxy to PSF size ratio peaks at lower values than the
ones that are used in the input simulation but there is a tail to larger values.
Overall the mean is biased low by around 10\%.
The flux ratio is skewed to larger values and overestimated by around 10\%.
Moreover, this distribution has two modes; one close to the truth, and one close to $F_b/(F_b+F_d)=1.5$.

The shear measurement biases thus depend on the galaxy intrinsic ellipticity in a non-trivial way. 
However, this can be converted into the shear measurement bias for a population of galaxies at different orientations using the ring test.
This is discussed in greater detail in Appendix \ref{Appendix1}.
We effectively perform a ring-test to obtain the shear calibration metrics described in Section~\ref{sec:defns_systematics}.

For the default galaxy and image parameters we find a multiplicative shear measurement bias of a few per cent. 
For an intrinsic galaxy ellipticity of 0.3 we find $m=0.03$ which is an order of magnitude larger than the requirement for upcoming surveys. 
The additive shear measurement bias is around $c=2\times 10^{-3}$ which is larger than the requirement for upcoming surveys, and around an order of magnitude larger than the requirement for far-future surveys.

\begin{center}
\begin{figure*}
\epsfig{file=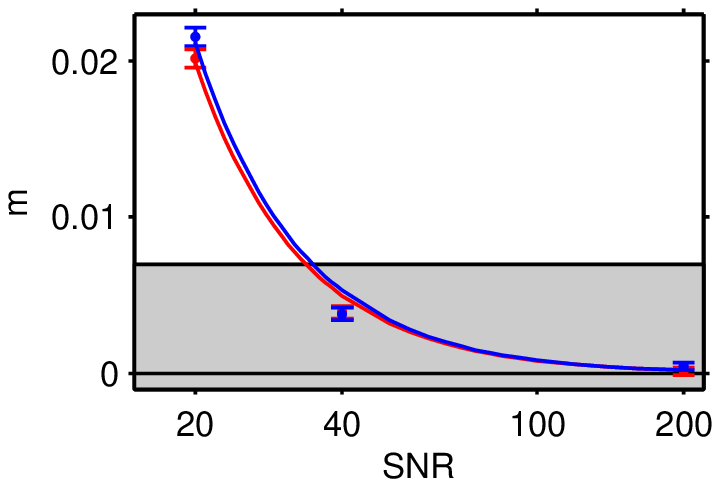,width=8cm}
\epsfig{file=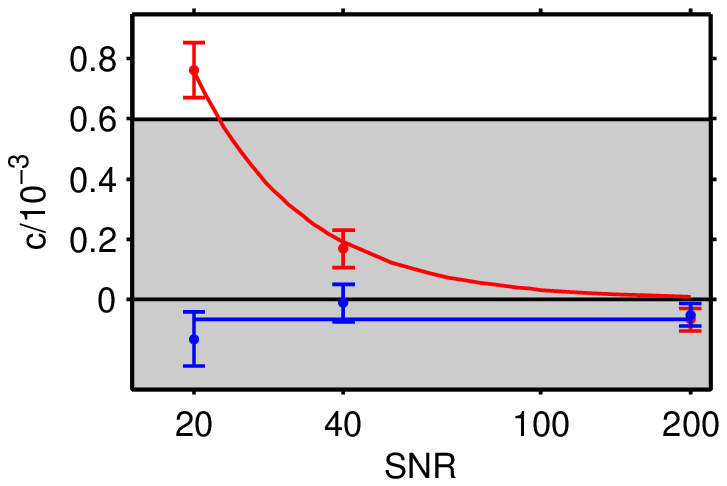,width=8cm}
\epsfig{file=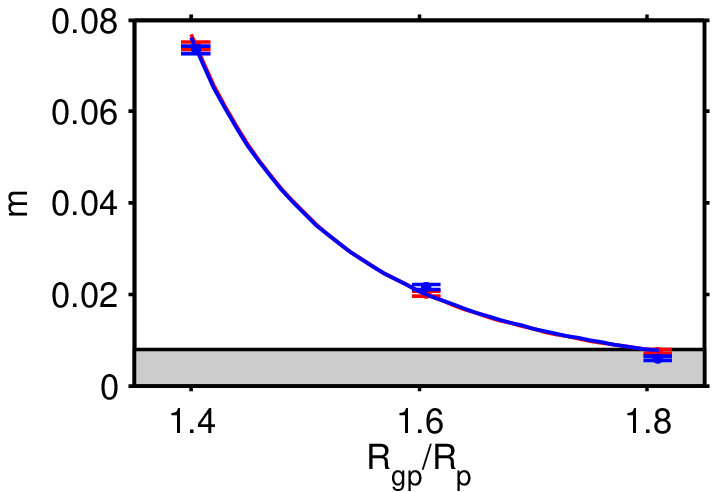,width=8cm}
\epsfig{file=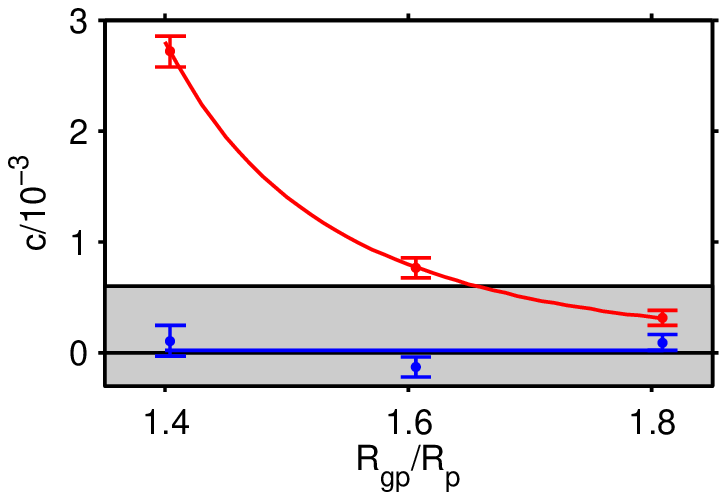,width=8cm}
\epsfig{file=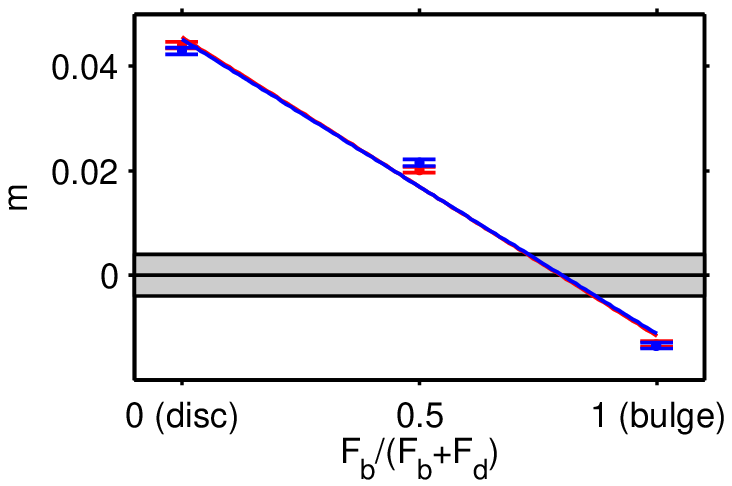,width=8cm}
\epsfig{file=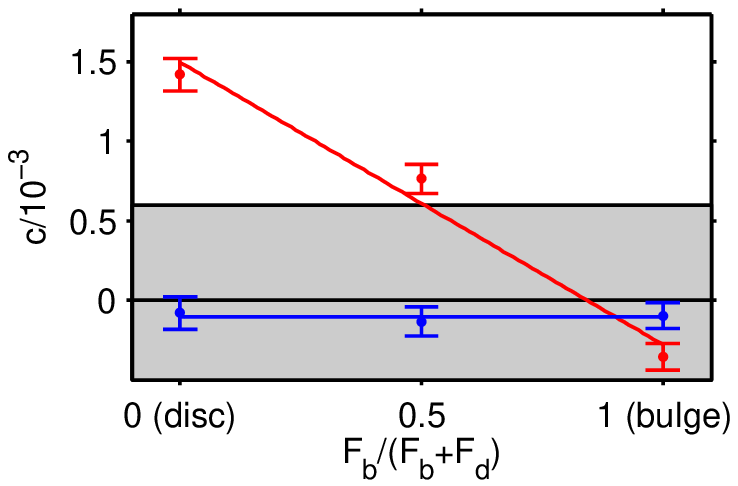,width=8cm}
\epsfig{file=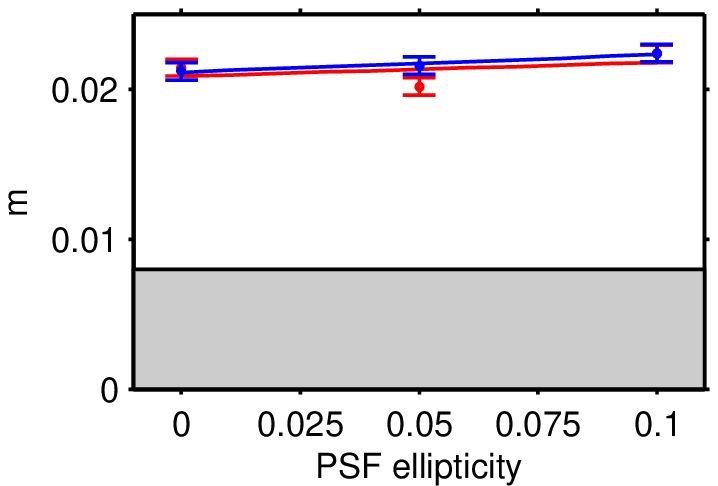,width=8cm}
\epsfig{file=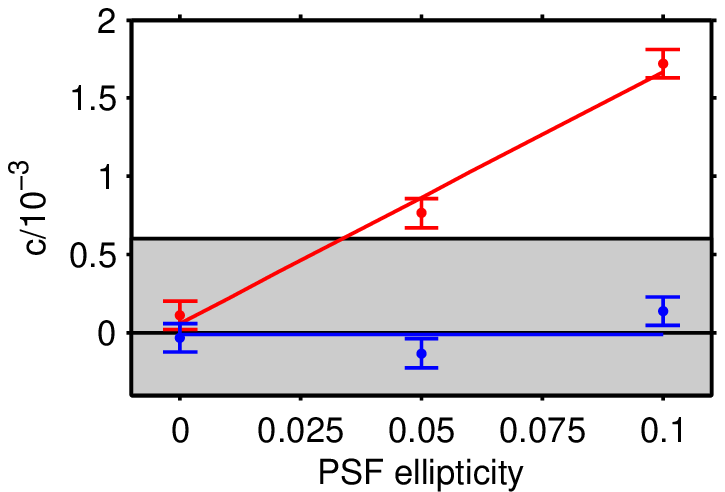,width=8cm}
\caption{Multiplicative (left column) and additive (right column) bias as a function of galaxy and image parameters at intrinsic ellipticity of $|e|=0.3$. 
First and second ellipticity components are marked with red and blue, respectively. Note that on some of the plots the errorbars are too small to be visible. 
Typical standard error on the multiplicative bias was of order (5 - 10) \e{-4} and on additive bias of order (5 - 10) \e{-5}. 
Lines are fits to the measured points, not the theoretical prediction. 
$m_1$, $m_2$ and $c_1$ as a function of SNR were fitted with $\mathrm{SNR}^{-2}$ function, $c_2$ with a constant. 
For $m_1$, $m_2$ and $c_1$ vs $R_{gp}/R_{p}$ the basis expansion for the fit was $\{(R_{gp}/R_{p})^{-2}, (R_{gp}/R_{p})^{-3}\}$, for $c_2$ - a constant. 
For the other parameters a linear fit was used. 
Appendix \ref{Appendix2} contains the data points (Table \ref{tab:DataPoints}) and equations for fitted functions (Table \ref{tab:FitParams}). 
The grey shaded area corresponds to requirements for upcoming surveys.
}
\label{fig:NbcModel} 
\end{figure*}
\end{center}

The multiplicative and additive shear measurement bias is shown as a function of galaxy and image parameters in Figure~\ref{fig:NbcModel}. 
Data points for those plots are listed in Table \ref{tab:DataPoints}, and the functions we fitted are given in equations in Table \ref{tab:FitParams}, both in Appendix \ref{Appendix2}.

The upper panels show the dependence on the image SNR.
This demonstrates clearly that the bias we observe is truly a noise bias, since the biases tend to zero at high SNR. 
Indeed for a SNR of 200 the biases are well below the requirement even for far-future surveys.
The dependence on SNR is well described by a quadratic function, shown
as a fitted line, as discussed anecdotally (Bernstein, priv. com.) and as expected from the derivations in \citet{hirataetal04} and R12. 
                                        
The upper middle panels of Figure~\ref{fig:NbcModel} show the dependence on the ratio of convolved galaxy to PSF size, as defined in Section~\ref{sec:defaults}.
The derivations in R12 showed that for Gaussian functions, the bias on the size parameter increases with the size of the PSF (Eq. 17). 
In our simulations the bias on the shear has a similar trend, as we observe an increased bias with decreased galaxy size relative to the PSF.
The bias is reduced by a factor of almost three when the convolved galaxy to PSF size increases from 1.41 to the default value of 1.62.
We modelled this dependence by using inverse power expansion with terms in $(R_{gp}/R{p} - 1)^{-2}$ and $(R_{gp}/R{p} - 1)^{-3}$.

The lower middle panels of Figure~\ref{fig:NbcModel} show the bias as a function of the flux ratio. 
Both multiplicative and additive bias change signs when the galaxy light profile changes from bulge to disc. 
Bulges are underestimated and discs are overestimated. 
This peculiar behaviour of the bias demonstrates the complexity of this problem. 
We use a straight line to fit the points, and this works reasonably well.  
 
The dependence on PSF ellipticity is shown in the bottom panels of  Figure~\ref{fig:NbcModel}.
As expected, e.g.\ from \citet{PaulinHenrikkson2008}, the dependence of the additive shear measurement bias is much greater than that of the multiplicative bias.
The additive shear bias dependence is very close to linear (shown by the fitted lines). 
Rotational symmetries in the problem, also visible on Figure \ref{fig:polyfit} indicate that there is very little dependence on the pixel orientation with respect to the PSF and galaxy. 
This essentially means that we can use results for the PSF aligned with the x - axis for any other PSF angle, by rotating the coordinate system.   

%% file: nbc_paper_calibration.tex
\section{Noise bias calibration} %S1
\label{sec:Application}

In this section we investigate how the bias measurements can be used to calibrate out the noise bias effect. 
First, we create a model of the bias on the ellipticity measurement as a function of four measured parameters: 
$\hat e_1$ , $\hat e_2$ , $\widehat{Rgp/Rp}$ , $\widehat{F_b/(F_b+F_d)}$, 
similar to Figure \ref{fig:polyfit} (note that we do not directly use the functions presented on Figure \ref{fig:NbcModel}, as they show a bias on shear in the form of $m$ and $c$, instead of the bias on the ellipticity).
We apply an additive correction predicted by our model directly to the measured ellipticity values. 
Finally we verify the accuracy of this procedure by testing it using a ring test consisting of 10 million noisy fiducial galaxies. 

This approach will not provide a perfect calibration, as our model of biases is calculated for a set of galaxies with particular true galaxy and image properties.
In practice we will only know the measured galaxy parameters, which are noisy, as illustrated in Figure \ref{fig:histograms}.
Therefore, if we read off the bias values from the measurements of the noisy measured galaxy parameters they will not be exactly the correct bias values for that galaxy. 
In this section, we investigate the scale of this effect.

The estimator of the ellipticity $\hat e$ is biased, so that $\hat e = \tilde e + b[\hat e]$, where $\tilde e$ is the unbiased estimator. 
By definition $\tilde  e$ averaged over noise realisations is equal to the true ellipticity, so that $\mean{\tilde e} = e^{\rm true}$.

We estimate the true shear $g$ with an estimator $\hat g$ in a ring test.
We write the following equations to show mathematically what is happening when we do the correction on the individual galaxy ellipticities. 
\begin{align}
 \hat g = \mean{\mean{  \widehat{e^{l}} }_N}_R &= \mean{   \mean{\widetilde{e^{l}}}_N + b[\widehat{e^{l}}]   }_R \\ 
&= g + \mean{ b[\widehat{e^{l}}] }_R
\label{eqn:shear_estimator}  
\end{align}
where $e^{l}=e+g$ is the lensed ellipticity, and subscripts $N$ and $R$ denote averages over noise realisations and around the ring respectively. 
Eq. \ref{eqn:shear_estimator} shows that the bias of the shear estimator will be equal to the bias on the lensed ellipticity $e+g$, averaged over noise realisations and the ring. 
This is the bias we aim to calibrate.

We create a correction model which describes $b[\hat e]$ as a function
of four galaxy parameters, i.e.
\begin{align}
 b[\hat e] = \beta(\theta) = \beta(e_1,e_2,R_{gp}/R_{p},F_b/(F_b+F_d))
\label{eqn:be_model}
\end{align}
Then we apply this correction to the noisy estimates $\hat \theta$, creating an estimator of the correction $\beta(\hat \theta)$ and we update our ellipticity estimate to be
\begin{align}
\hat e^\beta \leftarrow \hat e - \beta(
%S4\hat e,\dots).
%S4 \hat e_1,\dots).
% TK I would keep the dots because they indicate that in the model there is more parameters than the ellipticity
\hat \theta).
\label{eqn:correction}
\end{align}
Using this correction in the ring test implies
\begin{align}
\hat g^\beta  = g + \mean{ b[\widehat{e + g}] - \mean{\beta(
%S4 \hat e,\dots)
\hat \theta) }_N }_R .
\end{align}
Because we are applying the correction to the noisy maximum likelihood
estimates, the correction itself can be biased under noise, so that 
$b[\beta(\hat e,\dots)] = \mean{\beta(
%S4\hat e,\dots)
\hat \theta) } - b[\hat e]$.
Including this `bias on the correction', we expect the the final bias on the shear after applying our calibration procedure to be 
\begin{align}
b[\hat g^\beta] &= \mean{\mean{b[\beta(\widehat{e+g})]}_N}_R \\
\label{eqn:bias_on_calibrated_shear}
c^\beta &= \mean{\mean{b[\beta(\hat e)]}_N}_R \\
m^\beta &= \frac{\mean{\mean{b[\beta(\widehat{e +g})]}_N}_R - \mean{\mean{b[\beta(\hat e)]}_N}_R}{g}  .
\end{align}
Testing this procedure will include finding out how big the term in Eq. \ref{eqn:bias_on_calibrated_shear} is. 

In practice we create the model of the bias $\beta(\theta)$ (Eq. \ref{eqn:be_model}) using a learning algorithm based on Radial Basis Functions (RBF) Interpolation
\footnote{http://www.mathworks.com/matlabcentral/fileexchange/10056}, 
trained on all our simulated results. 
Then we use Eq. \ref{eqn:correction} to correct the ellipticity estimates.

The calibration procedure was tested by generating nearly ten million galaxy images using the default galaxy parameters. 
The ring test was performed as follows:
a set of galaxies was simulated with the galaxy intrinsic ellipticity angles equally spaced at 16 values from 0 to $\pi$,
(i)  with no shear applied
(ii) with a shear of $g_1 = 0.1$ applied.
In total 300,000 galaxies were simulated at each angle in the ring, for each shear value.
To compute the uncalibrated shear measurement bias, the measured ellipticity was averaged over all galaxies with a given shear to obtain a shear estimate for that population.
Then a straight line was fitted to the resulting shear estimates as a function of input shear to obtain the usual $m$ and $c$. 
To compute the calibrated shear measurement bias, the measured ellipticities were corrected using Eq. \ref{eqn:correction} before averaging to obtain the shear estimate. 

\begin{center}
\begin{figure}
\epsfig{file=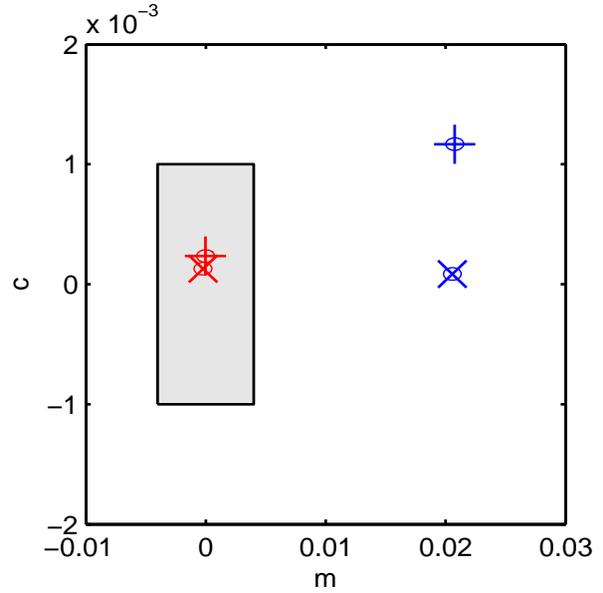,width=8cm,height=8cm} 
\caption{Values of multiplicative ($m$) and additive ($c$) bias for uncalibrated (blue) and calibrated (red) shear estimates.
Ellipses indicate one sigma error bars.
}
\label{fig:CalibrationTestResults} 
\end{figure}
\end{center}

The uncalibrated and calibrated shear measurement biases are presented in Figure \ref{fig:CalibrationTestResults}. 
We see that the uncalibrated shear measurement biases are well outside the requirement for upcoming surveys, as discussed earlier.
The calibration reduces the additive bias by a factor of around three, and the multiplicative bias by a factor of around ten.
We find that the bias term in Eq. \ref{eqn:bias_on_calibrated_shear} is insignificantly small to the accuracy afforded by our simulations.
%S4 [I guess handy if we can say if we did it to the accuracy of far-future surveys or not, or whether the number of sims limits us from drawing conclusions there]
% I was thiniking about it, because std_m2 = 0.000737 std_c2 = 0.000052 so it should be fine
Therefore the calibrated biases are now within the requirement for upcoming surveys for both additive and multiplicative shear biases.

%% file: nbc_paper_summary.tex
\section{Conclusions}
\label{sec:Wrap}

%S3 [noting some things to discuss from earlier in the text:
%S3 * model bias and noise bias - would our calibration method have worked with more free parameters?
%S3 * would ourm ethod have worked with a different PSF?]
%S3 * explain why there are not more points on the plots - done
%S4 [!!! we should summarise what was done in hirataetal04 somewhere]

In this paper we have investigated the effect of noise on shear
measurement from galaxy images. 
We have found that this can significantly bias shear measurement from realistic images, even though the bias goes away completely for images with lower noise levels.
This was previously studied in \citep{hirataetal04} and R12, who demonstrated the existence of this noise bias effect.
We quantified noise bias using images simulated from more realistic galaxy models and we used a forward fitting shear measurement method which fitted a matching set of galaxy models to the simulations (\textsc{Im3shape}, Zuntz et al. in prep).
These models are based on observationally-motivated combinations of exponential disk and de Vaucouleurs bulge models and are broadly representative of the light profiles of realistic galaxies. They have also formed the basis of previous weak lensing simulation programmes \citep{step1,great08results,great10results}.
We use a maximum likelihood estimator (MLE) to obtain galaxy ellipticity estimates from the images, and use these ellipticity estimates as our noisy shear estimates. 
We find that the shear measurement biases often exceed $\sim$1$\%$ and even approach $\sim$10$\%$ for the smallest galaxies and highest noise values we consider in this paper. 

One feature of the simulations presented is that they are deliberately internal: test galaxies are generated using the same models and routines used later for fitting them, the only difference being the addition of noise.  
In this way we are able to explore the effects of noise biases in isolation from the contribution of underfitting or model bias (e.g.\ \citealp{melchior2010,voigt2010,bernstein2010}).  
The fact that the biases we detect are considerable, even when fitting with perfect knowledge of the parametric galaxy model, is striking.  
We conclude that, for many methods, bias from unavoidable noise in galaxy images must be considered an important potential source of systematic error when seeking shear inference at sub-percent level accuracy.
The existence of noise bias is likely to be a common feature to many shape measurement methods (\citealp{hirataetal04}; R12).
Unless shape measurement methods are theoretically constructed to avoid noise bias, empirical calibration with simulations is necessary.

We quantified the noise bias as a function of image and galaxy parameters and found a strong dependence.
We found that the dependence on image signal-to-noise ratio is inverse square, as expected from symmetry arguments (e.g. see R12).
The dependence on galaxy size is quite non-linear and rises steeply as the galaxy size decreases relative to the PSF size. 
The bias depends on the galaxy profile in a complicated way. We find that for our fiducial parameters shears are overestimated for exponential disc galaxies and underestimated for de Vaucouleurs bulge galaxies. 
The dependence on bulge to total flux ratio is reasonably consistent with a linear relation.
There is a good linear relation between the additive shear measurement noise bias and the PSF ellipticity. 
%S4 [could add comment on dependence on galaxy intrinsic ellipticity]

%S2 [tidying up for JPL]
%S2 \textbf{***BARNEY TO DO: PARAGRAPH OR TWO DESCRIBING SOME OF THE OUTSTANDING QUESTIONS THAT COME FROM THE DETAILED RESULTS WHEN AVAILABLE?***}

Many shape measurement methods are potentially subject to noise bias, and for these methods this sort of calibration will be an important step in order to reduce systematic errors below the level required for upcoming survey datasets.  
We illustrate a correction scheme based on a model of the measured biases, as function of observed galaxy properties. 
%S4 [need to pre-empt more with why this is not going to be a trivial exercise. Added the followign sentence:]
%Note that this is not guaranteed to work because the observed galaxy properties are not the true galaxy properties and therefore we will be using slightly the wrong bias correction. %S4
%TK  This sounds a little pessimistic and may give a wrong message. We can actually measure how well we expect it to work. Reworded
Note that this is not expected to remove the bias completely because the observed galaxy properties are not the true galaxy properties and therefore we will be using slightly the wrong bias correction. %S4
This correction was able to reduce ellipticity estimator biases to
lower levels than those required for the upcoming lensing surveys, for a fiducial galaxy with SNR=20 and a typical intrinsic ellipticity of magnitude 0.3.  

There is a small residual bias remaining after this first level of correction. 
This is due to the scatter and bias in measured galaxy parameters about their true values. 
This scatter and bias is an output of the simulations and could therefore be propagated into a second level of bias correction which would reduce the residual bias yet further, into the realm of far-future surveys.

The calibration scheme we proposed can only be applied to a method which, in addition to ellipticity, also produces estimates of other parameters; it will probably be difficult to use it with a method such as KSB, which primarily aims to estimate only the ellipticity parameters.
 
This calibration approach is extremely computationally expensive and would ideally be carried out for a large range and sampling of image and galaxy parameters. The resolution of our results was limited by the available computing time. 
The final results shown in this paper took over 1 year of CPU time.

These results use a simple galaxy model in both the simulations and the fit. 
In practice it will be necessary to investigate more complicated galaxy models for both.
However, the presented results are encouraging. %S4
%S4 However, measuring the biases required for such a correction model requires data: these can either come from simulations, as presented here, or from deep imaging of the real sky.   [I guess obvious that the biases have to be calculated somehow?]
For future surveys the simulated data must be carefully constructed in order to recreate realistic observing conditions, and the realistic properties of the underlying galaxies (the latter requirement poses greater difficulties than the former).  
The deep imaging of the real sky is potentially an expensive overhead for future surveys, but may prove necessary for confidence in the final results.  
%S4 Ideally, data of both types would be available for calibrating lensing estimators that in many cases are generically biased due to noise.  
Accurate estimates of gravitational shear from methods affected by noise bias will rely on consistent strategies for measuring and correcting these systematic effects.

The presented calibration scheme does not use the information about the galaxy parameters distribution in the universe. 
We found that the measured galaxy parameters were a sufficiently good proxy for the true galaxy parameters that the noise bias could be corrected well enough for upcoming surveys. If this result were generally true then this places less stringent requirements on the simulations because the galaxy population demographics would not need to match exactly with reality, and the simulations would only have to span a realistic range of galaxy parameters. %S4
However, different calibration schemes could be created based on the distributions of galaxy parameters.
The simplest solution would be to calculate one $m$ and $c$ for the whole population of galaxies, randomly drawing not only noise maps but also galaxy and image parameters from histograms of measured parameters from galaxies in the survey. 
Using this method is not limited to maximum likelihood fitting; potentially all shear measurements methods could be calibrated that way.

We have used a white Gaussian noise model. In general it should be possible to repeat this procedure for a case of correlated noise. 
It should also be possible to repeat the procedure for Poisson noise.
Our bias results will also depend on the number of parameters used in the fitting. 
We have used seven free parameters and fixed the ratio of radii of the bulge and disc galaxy components to unity. 
We also assumed no constant background in the image, whereas this could also be included as a free parameter in the fit. 
An uncertain variable background level would complicate the analysis further.

Another approach would be to use a fully Bayesian analysis: 
use the full likelihood distribution (or samples) of ellipticity given the noisy images and propagate this uncertainty to the cosmological parameters.
%S4 [I think we have to add a lensfit ref here really...]
%TK I would rather not add it because that is not what lensfit is doing
%\citep[see][]{lensfit1}.
In this case the calibration would not be necessary.

\section*{Acknowledgements}
TK, SB, MH, BR and JZ acknowledge support from the European Research Council in the form of a Starting Grant with number 240672. 
Part of BR's work was done at the Jet Propulsion Laboratory, California Institute of Technology, under contract with NASA.
We thank Gary Bernstein for suggesting the calibration approach and for many fruitful discussions.
The authors acknowledge the use of the UCL Legion High Performance Computing Facility, and associated support services, in the completion of this work. 
We thank Dugan Witherick for help with Legion Cluster.
We thank Cris Sabiu and Caroline Pung for helpful discussions.

%% file: nbc_paper_appendix1.tex
\appendix

\section{Measurement of the bias on the shear}
\label{Appendix1}

The multiplicative and additive bias was measured using the following procedure.

\begin{center}
\begin{figure*}
\epsfig{file=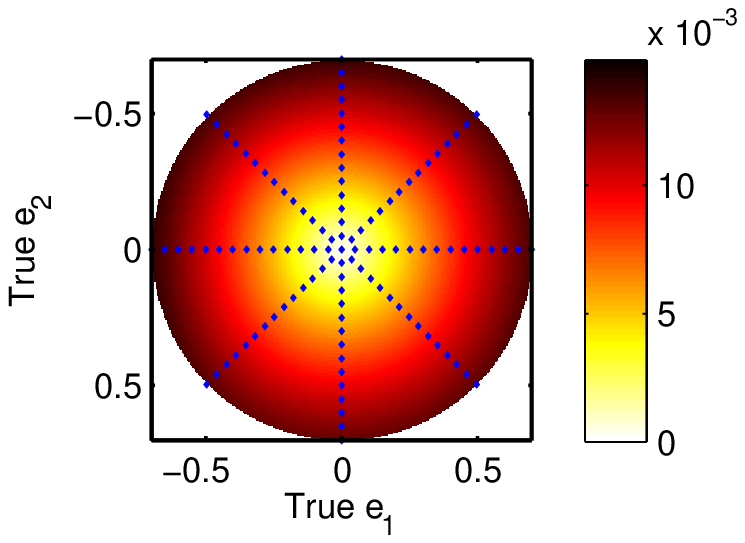,width=7cm}
\epsfig{file=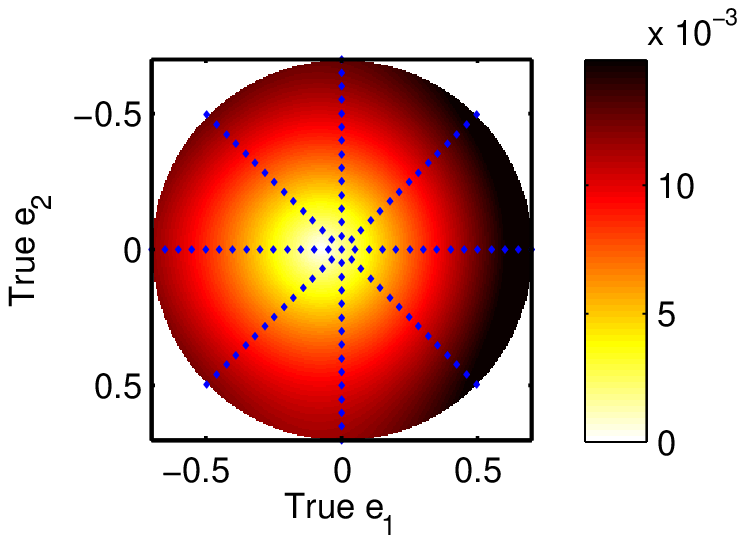,width=7cm}
\epsfig{file=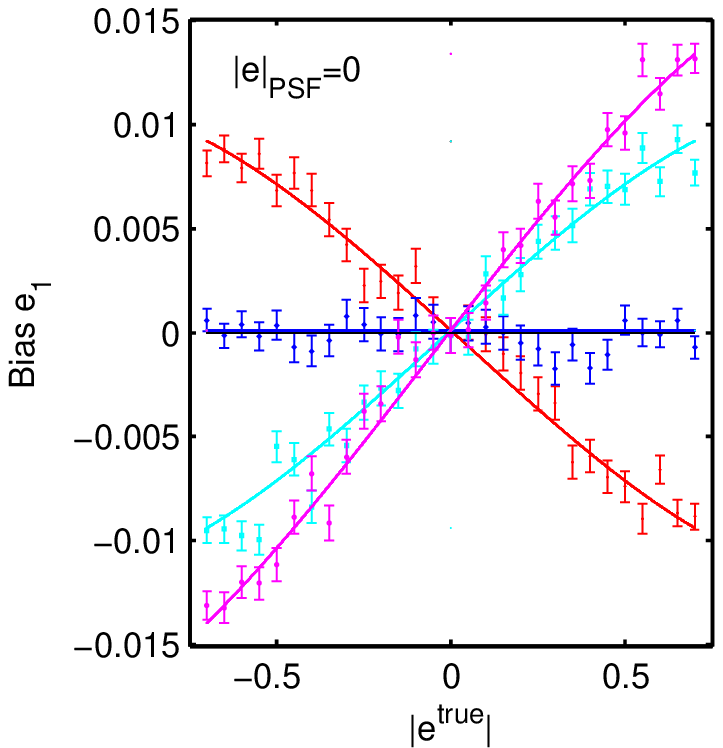,width=7cm}
\epsfig{file=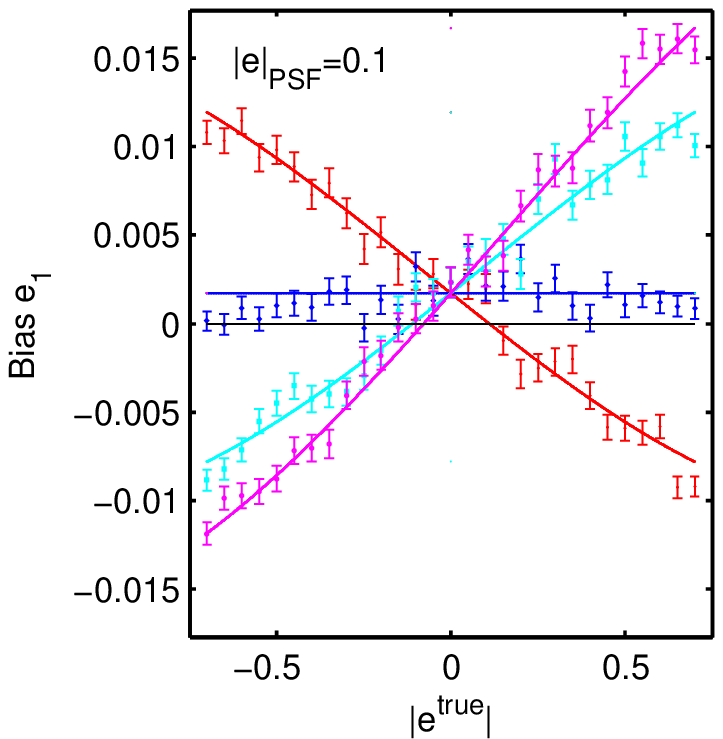,width=7cm} 
\epsfig{file=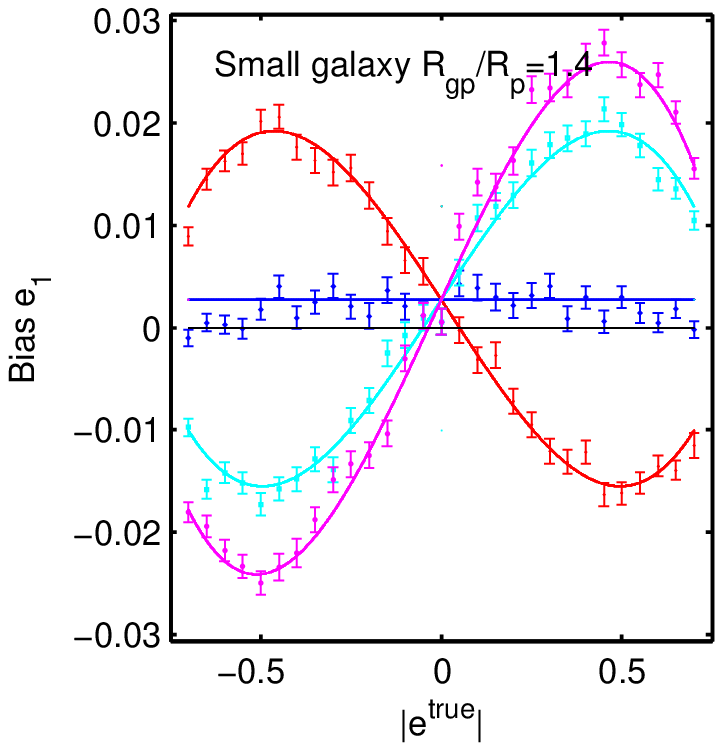,width=7cm}
\epsfig{file=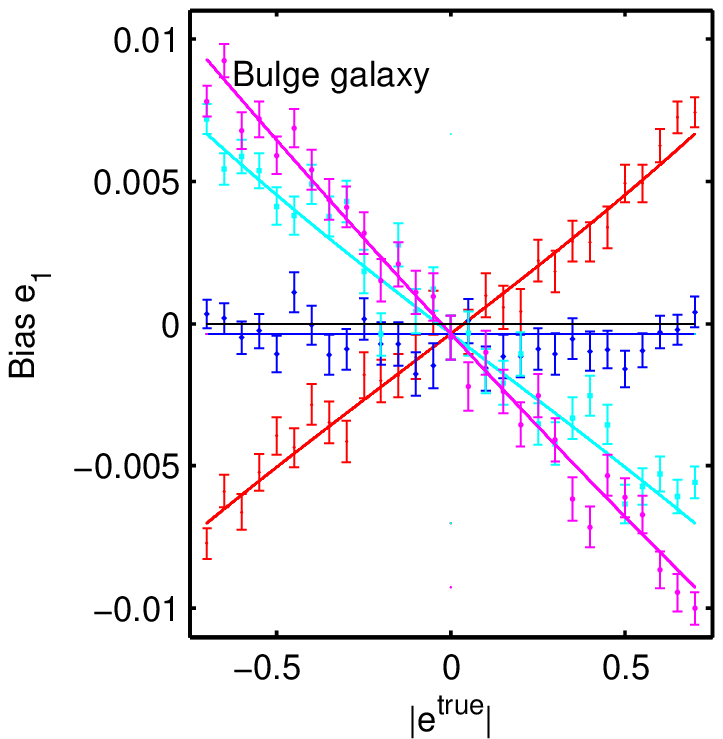,width=7cm}
\caption{
Top panels: bias on $\hat e_1$ as a function of true galaxy
ellipticity $e_1$ component. Colours (magenta, cyan, blue, red)
correspond to true ellipticity angles $\{0,\pi/8,\pi/4,3\pi/8\}$
joined with $\{\pi/2, 5\pi/8, 3\pi/4, 7\pi/8\}$. Lines are the third order polynomial fits to the points. 
Middle panels: colourscale presents the bias on $\hat e_1$ as a function of true galaxy ellipticity $e_1$ and $e_2$.
Ellipticity of the PSF was $e_{\rm PSF} = \{0.0,0.0\}$ for the left panel and $e_{\rm PSF} = \{0.1,0.0\}$ for the right panel.
Bottom left panel: galaxy with $R_{gp}/R_{p} = 1.4$. 
Notice that the bias on ellipticity changes sign for large intrinsic
ellipticities. This is due to the edge effect of the ellipticity parameter space, described in \ref{sec:Evaluation}. 
Bottom right panel: pure bulge galaxy.
Note that the bias has a different sign to the fiducial galaxy.   
}
\label{fig:polyfit} 
\end{figure*}
\end{center}

\begin{enumerate}
 \item {\bf Evaluate the bias on a grid in observed ellipticity:}
A grid in observed ellipticity parameter was created for each test galaxy in Table \ref{tab:SimParams}. This grid consisted of 8 angles on a ring. At each angle, 15 ellipticity magnitudes were used in range $\{0, 0.05, \dots , 0.7\}$. This grid is presented in Figure \ref{fig:polyfit}. For each point on this grid, we evaluate 20000 noise realisations, and average them to obtain the bias. The number of noise realisations is chosen so that the uncertainty on the mean was smaller than $\sigma_e < 10^{-3}$.
\item {\bf Create a model of the bias as a function of observed ellipticity:}
A third order 2D polynomial was fit to the surface of the bias. Not
all terms in the 2D expansion were used to avoid overfitting of the
data. In particular, we used $\{1,e_1,e_1^2,e_2^2,e_1^3\}$ for fitting
the bias on $e_1$, analogously for $e_2$. This expansion takes into
account the inherent rotational symmetry of the problem: rotating galaxy ellipticity and PSF ellipticity vectors results in the rotation of the bias vector. 
\item {\bf Perform a ring-test to calculate $m$ and $c$:}
The parametric model of the bias surface allows us to perform a ring test at any desired intrinsic ellipticity.
\end{enumerate}
 
The bottom panels of Figures \ref{fig:polyfit} present the grid (dots)
and interpolated surface (colour scale) of the magnitude of bias as a function of true $e_1$ and $e_2$ for a circular and
elliptical PSF. We note that for circular PSF within the modelled range, the bias surface has a circular symmetry 
which demonstrates that the problem is symmetric and that the effect of the pixel orientation with
respect to the galaxy is not strong.  %  made this strong statement a bit less strong
The top panels of Figure \ref{fig:polyfit} present cross sections of the above grid and surface for each angle. 

\section{Parameters and functions used to create models of the bias on ellipticity and shear}
\label{Appendix2}

\begin{table*}
\center
\begin{tabular}{|l|c|c|c|c|}
\hline
		& $m_1$ & $m_1$ & $c_1$ & $c_2$  \\
\hline
fiducial &$ +0.01962 \pm   0.00040 $&$ +0.02094 \pm  0.00044 $&$ +0.00084 \pm  0.00040 $&$ -0.00006 \pm  0.00008 $ \\ 
 $\textrm{SNR}=200$ &$ +0.00010 \pm   0.00017 $&$ +0.00021 \pm  0.00018 $&$ -0.00007 \pm  0.00017 $&$ -0.00003 \pm  0.00004 $ \\ 
 $\textrm{SNR}=40$ &$ +0.00383 \pm   0.00028 $&$ +0.00416 \pm  0.00031 $&$ +0.00016 \pm  0.00028 $&$ -0.00009 \pm  0.00006 $ \\ 
 $R_{gp}/R_{p}=1.4$ &$ +0.05809 \pm   0.00060 $&$ +0.05263 \pm  0.00067 $&$ +0.00274 \pm  0.00060 $&$ -0.00003 \pm  0.00013 $ \\ 
 $R_{gp}/R_{p}=1.8$ &$ +0.00809 \pm   0.00031 $&$ +0.00688 \pm  0.00034 $&$ +0.00037 \pm  0.00031 $&$ +0.00009 \pm  0.00007 $ \\ 
 $\textrm{disc}$ &$ +0.03992 \pm   0.00045 $&$ +0.03929 \pm  0.00050 $&$ +0.00166 \pm  0.00045 $&$ -0.00002 \pm  0.00010 $ \\ 
 $\textrm{bulge}$ &$ -0.01325 \pm   0.00036 $&$ -0.01171 \pm  0.00040 $&$ -0.00036 \pm  0.00036 $&$ -0.00006 \pm  0.00008 $ \\ 
 $e_{\rm PSF}=\{0.0,0.0\}$ &$ +0.02050 \pm   0.00040 $&$ +0.02067 \pm  0.00044 $&$ +0.00010 \pm  0.00040 $&$ -0.00005 \pm  0.00008 $ \\ 
 $e_{\rm PSF}=\{0.1,0.0\}$ &$ +0.02176 \pm   0.00040 $&$ +0.02119 \pm  0.00044 $&$ +0.00195 \pm  0.00040 $&$ +0.00006 \pm  0.00009 $ \\ 
 $R_{gp}/R_{p}=1.8 \ \textrm{disc}$ &$ +0.01740 \pm   0.00035 $&$ +0.01785 \pm  0.00039 $&$ +0.00089 \pm  0.00035 $&$ -0.00006 \pm  0.00007 $ \\ 
 $R_{gp}/R_{p}=1.8 \ \textrm{bulge}$ &$ -0.02694 \pm   0.00031 $&$ -0.02101 \pm  0.00034 $&$ -0.00066 \pm  0.00031 $&$ -0.00012 \pm  0.00007 $ \\ 
 $R_{gp}/R_{p}=1.4 \ \textrm{disc}$ &$ +0.05899 \pm   0.00063 $&$ +0.05634 \pm  0.00070 $&$ +0.00289 \pm  0.00063 $&$ -0.00003 \pm  0.00013 $ \\ 
 $R_{gp}/R_{p}=1.4 \ \textrm{bulge}$ &$ +0.03450 \pm   0.00054 $&$ +0.03459 \pm  0.00060 $&$ +0.00161 \pm  0.00054 $&$ +0.00016 \pm  0.00012 $ \\ 
\hline
\end{tabular}
\caption{Measured multiplicative and additive biases for all simulated galaxies. Biases here are shown for a ring test using intrinsic ellipticity of 0.3. 
All parameters of the galaxies were the same as the fiducial model, except the ones indicated in the first column.}
\label{tab:DataPoints}
\end{table*}

\begin{table*}
\center
\begin{tabular}{|l|r|r|}
\hline
		& $m_1$ & $c_1$    \\
		& $m_2$ & $c_2$    \\
\hline
$D_1$ := $\textrm{SNR}$  & $+7.956\e{+00} \cdot D_1^{-2}$  & $+3.026\e{-01} \cdot D_1^{-2}$  \\  
 & $+8.470\e{+00} \cdot D_1^{-2}$  & $-6.685\e{-05} $  \\ \hline 
$D_2$ := $R_{gp}/R_{p} - 1$  & $-2.190\e{-03} \cdot (D_2)^{-2} +5.791\e{-03} \cdot (D_2)^{-3}$  & $-4.002\e{-05} \cdot (D_2)^{-2} +1.953\e{-04} \cdot (D_2)^{-3}$  \\  
 & $-1.923\e{-03} \cdot (D_2)^{-2} +5.639\e{-03} \cdot (D_2)^{-3}$  & $+2.089\e{-05} $  \\ \hline 
$D_3$ := $\frac{F_b}{F_b+F_d}$  & $-5.716\e{-02} +4.557\e{-02}\cdot D_3 $  & $-1.775\e{-03} +1.496\e{-03}\cdot D_3 $  \\  
 & $-5.641\e{-02} +4.518\e{-02}\cdot D_3 $  & $-1.034\e{-04} $  \\ \hline 
$D_4$ := $e_{PSF}$  & $+2.084\e{-02} +9.193\e{-03}\cdot D_4 $  & $+5.697\e{-05} +1.612\e{-02}\cdot D_4 $  \\  
 & $+2.111\e{-02} +1.185\e{-02}\cdot D_4 $  & $-1.107\e{-05} $  \\ \hline 
\end{tabular}
\caption{Equations for noise bias model function. These are the equations fitted to the data points in Figure \ref{fig:NbcModel}.}
\label{tab:FitParams}
\end{table*}

\begin{table*}
\center
\begin{tabular}{|l|c|c|c|c|c|c|c|c|c|c|}
\hline
		& $a_1^{(0)}$ & $a_1^{(1)}$ & $a_1^{(2)}$ & $a_1^{(3)}$ & $a_1^{(4)}$ & $a_2^{(0)}$ & $a_2^{(1)}$ & $a_2^{(2)}$ & $a_2^{(3)}$ & $a_2^{(4)}$   \\
\hline

fiducial &$ +0.0008 $&$ +0.0201 $&$ +0.0003 $&$ +0.0004 $&$ -0.0013 $ &$ -0.0001 $&$ +0.0216 $&$ +0.0006 $&$ -0.0047 $&$ -0.0070 $ \\ 
 $\textrm{SNR}=200$ &$ -0.0001 $&$ +0.0001 $&$ +0.0004 $&$ -0.0009 $&$ +0.0004 $ &$ -0.0001 $&$ +0.0004 $&$ +0.0002 $&$ -0.0009 $&$ -0.0007 $ \\ 
 $\textrm{SNR}=40$ &$ +0.0002 $&$ +0.0039 $&$ +0.0004 $&$ -0.0013 $&$ +0.0013 $ &$ -0.0000 $&$ +0.0037 $&$ +0.0001 $&$ +0.0047 $&$ +0.0021 $ \\ 
 $R_{gp}/R_{p}=1.4$ &$ +0.0027 $&$ +0.0767 $&$ -0.0076 $&$ -0.1154 $&$ -0.1075 $ &$ +0.0001 $&$ +0.0754 $&$ -0.0031 $&$ -0.1048 $&$ -0.1125 $ \\ 
 $R_{gp}/R_{p}=1.8$ &$ +0.0003 $&$ +0.0073 $&$ +0.0008 $&$ +0.0055 $&$ +0.0054 $ &$ +0.0001 $&$ +0.0059 $&$ -0.0003 $&$ +0.0073 $&$ +0.0066 $ \\ 
 $\textrm{disc}$ &$ +0.0014 $&$ +0.0443 $&$ +0.0004 $&$ -0.0282 $&$ -0.0256 $ &$ -0.0001 $&$ +0.0432 $&$ +0.0004 $&$ -0.0247 $&$ -0.0254 $ \\ 
 $\textrm{bulge}$ &$ -0.0004 $&$ -0.0132 $&$ +0.0007 $&$ -0.0024 $&$ -0.0000 $ &$ -0.0001 $&$ -0.0137 $&$ +0.0002 $&$ -0.0023 $&$ +0.0133 $ \\ 
 $e_{PSF}=\{0.0,0.0\}$ &$ +0.0001 $&$ +0.0216 $&$ -0.0008 $&$ -0.0072 $&$ -0.0042 $ &$ -0.0000 $&$ +0.0213 $&$ -0.0009 $&$ -0.0046 $&$ -0.0048 $ \\ 
 $e_{PSF}=\{0.1,0.0\}$ &$ +0.0017 $&$ +0.0223 $&$ +0.0014 $&$ -0.0056 $&$ -0.0039 $ &$ +0.0001 $&$ +0.0227 $&$ -0.0015 $&$ -0.0006 $&$ -0.0107 $ \\ 
 $R_{gp}/R_{p}=1.8, \ \textrm{disc}$ &$ +0.0008 $&$ +0.0172 $&$ +0.0010 $&$ -0.0008 $&$ +0.0049 $ &$ -0.0000 $&$ +0.0173 $&$ -0.0002 $&$ +0.0052 $&$ +0.0016 $ \\ 
 $R_{gp}/R_{p}=1.8, \ \textrm{bulge}$ &$ -0.0006 $&$ -0.0290 $&$ +0.0003 $&$ +0.0030 $&$ +0.0140 $ &$ -0.0001 $&$ -0.0230 $&$ -0.0004 $&$ +0.0020 $&$ +0.0108 $ \\ 
 $R_{gp}/R_{p}=1.4, \ \textrm{disc}$ &$ +0.0030 $&$ +0.0843 $&$ -0.0110 $&$ -0.1420 $&$ -0.1485 $ &$ +0.0001 $&$ +0.0844 $&$ -0.0010 $&$ -0.1540 $&$ -0.1425 $ \\ 
 $R_{gp}/R_{p}=1.4, \ \textrm{bulge}$ &$ +0.0014 $&$ +0.0425 $&$ -0.0012 $&$ -0.0500 $&$ -0.0468 $ &$ +0.0001 $&$ +0.0450 $&$ -0.0004 $&$ -0.0593 $&$ -0.0529 $ \\  
\hline 
\end{tabular}
\caption{Parameters of equations for the bias on ellipticity. These are the parameters used with Eq. \ref{eqn:polynomial}. 
}
\label{tab:BiasEllipticityFitParams}
\end{table*}

Table \ref{tab:DataPoints} contains the multiplicative and additive bias measurements for all galaxies used in this work. 
See Appendix \ref{Appendix1} for details of how these values were calculated. 
Fiducial galaxy parameters were:
$\textrm{SNR}=20$,
$R_{gp}/R_{p}=1.6$,
$\mathrm{FWHM}_{\rm PSF}=2.85$,
$e_{\rm PSF}=\{0.05,0\}$,
$\beta^{\rm Moffat}=3$,
$\textrm{flux}_{\rm bulge}/\textrm{flux}_{\rm total}=0.5$,
$r_{\rm bulge}/r_{\rm disc}=1.0$. 
Table \ref{tab:FitParams} contains equations of the functions in Figure \ref{fig:NbcModel}.
Table \ref{tab:BiasEllipticityFitParams} contains the parameters of polynomial function fitted to the bias on ellipticity, for example in Figure \ref{fig:polyfit}.
The equation used with these parameters is
\begin{align}
b[\hat e_1] = a_1{(0)} + a_1^{(1)}\hat e_1 + a_1^{(2)}\hat e_1^2 + a_1^{(3)}\hat e_1^2 \hat e_2 +  a_1^{(4)}\hat e_1^3.
\label{eqn:polynomial} 
\end{align}
for $b[\hat e_1]$ accordingly with parameters $a_2$.